\documentclass[12pt]{iopart}
\usepackage{epsfig}

\usepackage{amssymb,amsmath}
\usepackage{graphics}
\usepackage{dcolumn}
\usepackage{bm}
\usepackage{epsfig}

\newcommand{\SEs}{{\rm self-energies}\ }
\newcommand{\NE}{{\rm non-equilibrium}\ }
\newcommand{\MB}{{\rm MB}}

\begin{document}
\jl{3}

\title[Non-equilibrium many-body transport]
{Non-equilibrium renormalised contacts for transport in nanodevices with interaction:
a quasi-particle approach}

\author{H. Ness, L. K. Dash}
\address{Department of Physics, University of York, Heslington, York YO10 5DD, UK}
\address{European Theoretical Spectroscopy Facility (ETSF)}

\begin{abstract}
  We present an application of a new formalism to treat the quantum
transport properties of fully interacting nanoscale junctions 
We consider a model single-molecule nanojunction in the presence of two kinds of 
electron-vibron interactions. In terms of the electron density matrix, one interaction is diagonal 
in the central region and the second off-diagonal between the central region
and the left electrode. We use a non-equilibrium Green's function technique to calculate
the system's properties in a self-consistent manner. The interaction self-energies are
calculated at the Hartree-Fock level in the central region and within a dynamical mean-field-like
approach for the crossing interaction.
Our calculations are performed for different transport regimes ranging from the far off-resonance
to the quasi-resonant regime, and for a wide range of parameters.
They show that a non-equilibrium (i.e. bias dependent) dynamical (i.e. energy dependent) 
renormalisation is obtained for the contact between the left electrode and the central region
in the form of a non-equilibrium renormalisation of the lead embedding potential.
The conductance is affected by the renormalisation of the contact: the amplitude of the
main resonance peak is modified as well as `the lineshape of the first vibron side-band.
\end{abstract}
 

\maketitle

\section{Introduction}
\label{sec:intro}


The theory of quantum electronic transport in nano-scale devices has evolved rapidly over 
the past decade,
as advances in experimental techniques have made it possible to measure transport 
properties down to single-molecule devices. 
The development of accurate theoretical methods for the description of quantum transport at 
the single-molecule level is essential for continued progress in a number of areas including 
molecular electronics, spintronics, and thermoelectrics \cite{Widawsky:2012}.

One of the longstanding problems in quantum charge transport is the establishment of a theoretical 
framework which allows for quantitatively accurate predictions of conductance from first principles. 
The need for methods going beyond the conventional approaches, based on equilibrium electronic
structure calculations combined  with Landauer-like elastic scattering, has been clear for a 
number of years. 
Only recently have more advanced methods to treat electronic interaction appeared (for example
see Refs.~\cite{Strange:2011,Rangel:2011,Darancet:2007}). 

Alternative frameworks to deal with the steady-state or time-dependent transport are given by 
many-body perturbation theory, based on the non-equilibrium (NE) Green's function (GF) formalism:
in these approaches, the interactions and (initial) correlations are taken into account by using 
conserving approximations for the many-body self-energy \cite{Baym:1962,vonBarth:2005,vanLeeuwen:2006,Kita:2010,Tran:2008,Myohanen:2008,Myohanen:2010,
Perfetto:2010,Velicky:2010,PuigvonFriesen:2009}.
 
Other interactions, such as electron-phonon coupling, also play an important role in 
single-molecule quantum transport. These interactions are more important in nanoscale 
systems, as the electronic probability density is concentrated in a small 
region of space and thus normal screening mechanisms are ineffective.
Such interactions are also crucial in inelastic electron tunneling spectroscopy.
Such a technique constitutes an important 
basis for spectroscopy of molecular junctions, yielding insight into the vibrational modes 
(single molecule phonons called vibrons) and ultimately the atomic structure of the 
junction \cite{Arroyo:2010}.

There have been many theoretical investigations focusing on the
effect of electron-vibron coupling in
molecular- and atomic-scale wires [16-51].
In all these studies, the interactions have always been considered to be present 
in the central region (i.e. the molecule) only, with the latter connected to two non-interacting
terminals. Interactions are also assumed not to cross at the contracts between the central region
and the leads. When electronic interactions are present throughout the system, as within 
density-functional theory calculations, they are treated only at the mean-field level and do not 
allow for any inelastic scattering events.
However, such approximations are valid only in a very limited number of practical cases. 
The interactions, in principle, exist throughout the entire system.

In a recent paper we derived a general expression for the current in nano-scale junctions
with interaction present everywhere in the system \cite{Ness:2011}.
The importance of such extended interactions in nano-scale devices has also been addressed, 
for electron-electron interaction, in recently developed approaches such as 
Refs.~\cite{Strange:2011,Perfetto:2012}.

In the present paper, we apply our formalism \cite{Ness:2011} to a specific model of single-molecule 
nanojunctions.
We focus on a model system in the presence of electron-vibron interaction both
within the molecule {\em and} between the molecule and one of the leads. 
We adopt a quasiparticle-like
approach to treat the crossing interactions (i.e. there are some restrictions on the components of
the self-energy for the crossing interaction).
We show how the interaction
crossing at one interface of the molecular nanojunctions affects the transport properties
by renormalising the coupling at the interface in a dynamical and bias-dependent manner.

The paper is organised as follows:
In Sec. \ref{sec:transport}, we briefly recall the main result of our expression for the current
in fully interacting systems.
In Sec. \ref{sec:interac}, we present the model Hamiltonian for the system which includes
two kinds of electron-vibron interaction: a Holstein-like Hamiltonian combined with a
Su-Schrieffer-Heeger-like Hamiltonian. In this section, we also describe how the corresponding
self-energies are calculated and the implications of such approximations on the current
expression at the left and right interfaces. 
In Sec. \ref{sec:res}, we show how the non-equilibrium dynamical renormalisation affects
the generalised embedding potential of the left ($L$) lead, and how in turn this affects the conductance
of the nanojunction.
We finally conclude and discuss extensions of the present work in Sec. \ref{sec:ccl}.

\section{General theory for quantum transport}
\label{sec:transport}

We consider a two-terminal device, made of three regions left-central-right, labelled $L-C-R$,
 in the steady-state regime.
In such a device the
interaction---which we specifically leave undefined
(e.g. electron-electron or electron-phonon)---is assumed to be
well described in terms of the single-particle self-energy $\Sigma^{\rm MB}$ and
spreads throughout the entire system. 

We use a compact notation for the Green's function $G$ and the self-energy $\Sigma$ 
matrix elements $M(\omega)$.  
They are annotated $M_C$ ($M_L$ or $M_R$) for the elements in the
central region $C$ (left $L$, right $R$ region respectively), and $M_{LC}$ (or $M_{CL}$)
and $M_{RC}$ (or $M_{CR}$) for the elements between region $C$ and region $L$ or $R$. 
There are no direct interactions between the two electrodes,
i.e. $\Sigma^{\rm MB}_{LR/RL}=0$.

In Refs.~\cite{Ness:2011,Ness:2012b}, we showed that for a finite applied bias $V$
the steady-state current $I_L(V)$ flowing through the left $LC$ interface is given by:
\begin{equation}
\label{eq:ILfinal}
\begin{split}
 I_L & = \frac{e}{\hbar} \int \frac{d\omega}{2\pi}
 {\rm Tr}_{\{C\}}\left[ G^r_C \tilde{\Upsilon}^{L,l}_C + G^a_C
  (\tilde{\Upsilon}^{L,l}_C)^\dagger  + G^<_C(\tilde\Upsilon^L_C -
  (\tilde{\Upsilon}^L_C)^\dagger) \right] \\
+ & {\rm Tr}_{\{L\}}\left[\Sigma^{\MB,>}_L G^<_L - \Sigma^{\MB,<}_L G^>_L \right]
\end{split}
\end{equation}
where the $\tilde\Upsilon_C^X$ quantities are 
\begin{equation}
\label{eq:Upsilons}
\begin{split}
\tilde\Upsilon^L_C(\omega)  & = \Sigma^a_{CL}(\omega)\ \tilde{g}^a_L(\omega)\ \Sigma^r_{LC}(\omega) , \\
(\tilde\Upsilon^L_C)^\dag & = \Sigma^a_{CL}\ \tilde{g}^r_L\ \Sigma^r_{LC} , \\
\tilde\Upsilon^{L,l}_C  & = \Sigma^<_{CL} \left( \tilde{g}^a_L - \tilde{g}^r_L \right) \Sigma^r_{LC}
+ \Sigma^r_{CL}\ \tilde{g}^<_L\ \Sigma^r_{LC} .
\end{split}
\end{equation}
By definition $\Sigma_{LC}(\omega)= V_{LC} + \Sigma^\MB_{LC}(\omega)$ (similarly for the $CL$
components) where $V_{LC/CL}$ are the nominal coupling matrix elements between the $L$ and $C$
regions.
$\tilde{g}^{x}_L(\omega)$ are the GF of the region $L$ renormalised by the interaction
{\em inside} that region, where $x=r,a,<$ stands for the retarded, advanced and lesser
GF components respectively. For example, for the advanded and retarded components, we have
$(\tilde{g}^{r/a}_L(\omega))^{-1} = ({g}^{r/a}_L(\omega))^{-1} - \Sigma^{{\rm MB},r/a}_L(\omega)$
where all quantities are defined only in the subspace $L$.

The second trace in Eq.~(\ref{eq:ILfinal}) corresponds to inelastic events induced 
by the interaction in the $L$ lead. At equilibrium, because of the detailed
balance equation $\Sigma^{>} G^< = \Sigma^{<} G^>$, this term vanishes.
At non-equilibrium, this is generally not the case.
However, when a local detailed balance equation holds, i.e. the system is locally
in a (quasi)equilibrium state, this terms vanishes since one recovers locally
$\Sigma^{\MB,>} G^< = \Sigma^{\MB,<} G^>$.
Hence, in practice, we do not have to worry about the infinite sum in the trace
${\rm Tr}_{\{L\}}[...]$ since there will always be a region/boundary in the left lead
$L$ beyond which the system is at (quasi)equilibrium.

The first trace in the current equation Eq.~(\ref{eq:ILfinal}) corresponds to a 
generalisation of the result of Meir and Wingreen \cite{Meir:1992}. It encompasses
the cases for which the interactions are present in the three $L, C, R$ regions as 
well as in between the $L/C$ and $C/R$ regions. 
It also bears some resemblance to the expression derived by Meir
and Wingreen \cite{Meir:1992} when written as:
\begin{equation}
\label{eq:IL_MeirWingreen_bis}
\begin{split}
I_L^{\rm MW} & = \frac{e}{\hbar} \int \frac{{\rm d}\omega}{2\pi} 
 {\rm Tr}_{\{C\}} \left[ G^r_C \Sigma^{L,<}_C + G^a_C (\Sigma^{L,<}_C)^\dag  
+ G^<_C  (\Sigma^{L,a}_C - \Sigma^{L,r}_C) \right] .
\end{split}
\end{equation}
where we use the standard definitions
$\Sigma^{L,<}_C = - (\Sigma^{L,<}_C)^\dag = V_{CL}\ {g}^<_L\ V_{LC} = {\rm i} f_L \Gamma_L$
and
$\Sigma^{L,a}_C - \Sigma^{L,r}_C = V_{CL} ({g}^a_L-g^r_L) V_{LC} = {\rm i}\Gamma_L$. 

By comparing Eq.~(\ref{eq:ILfinal}) and Eq.~(\ref{eq:IL_MeirWingreen_bis}), we can see
that the quantities $\tilde\Upsilon_{LC}$ ($\tilde\Upsilon^\dag_{LC}$)
and $\tilde\Upsilon^l_{LC}$ are now playing the role of the $L$ lead self-energy  
$\Sigma^a_L$ ($\Sigma^r_L$) and $\Sigma^<_L$ respectively
when the interactions cross at the $LC$ interface.

\section{Model for the interaction}
\label{sec:interac}

\subsection{Hamiltonians}
\label{sec:Hamiltonian}

We consider a single-molecule junction in the presence of electron-vibron interaction both
inside the central region and crossing at the contacts.
We further concentrate on a model for the central region which consists of a single molecular 
level coupled to a single vibrational mode.  
A full description of our methodology, for the interaction inside
the region $C$, is provided in Refs.~\cite{Dash:2010,Ness:2010,Dash:2011}.
Moreover, we consider that some electron-vibron interaction exists also at one
contact (the left $L$ electrode for instance).
This model typically corresponds to an experiment for a molecule chemisorbed onto a surface (the
left electrode) with a tunneling barrier to the right $R$ lead.

\begin{figure}
\centering  
  \includegraphics[clip,width=0.8\textwidth]{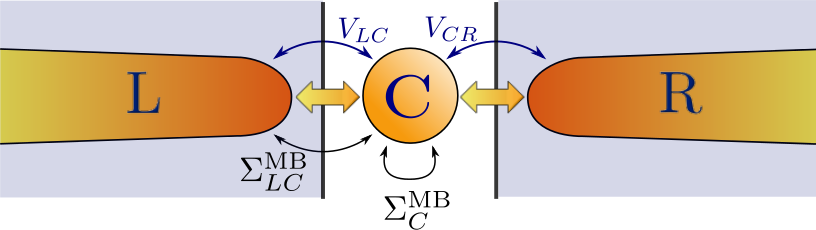}
  \caption{(Color online) Schematic representation of a central
scattering region $C$ connected to the left $L$ and right $R$ electrodes.
Interactions are given by the coupling of the region $C$ to the
$L$ ($R$) electrode ($V_{LC/CL}$ and $V_{RC/CR}$), and by the many-body effects 
within the central region ($\Sigma_C^{\rm MB}$) and at the $LC$ interface
($\Sigma_{LC}^{\rm MB}$). $\Sigma_C^{\rm MB}$ corresponds to the coupling of an
electron with the vibron mode $\omega_0$ (with coupling strength $\gamma_0$), and
$\Sigma_{LC}^{\rm MB}$ corresponds to the coupling of a hopping electron with another
vibron mode $\omega_A$ (with coupling strength $\gamma_A$).}
  \label{fig:schema}
\end{figure}

In the following model, we consider two kinds of electron-vibron coupling: 
a local coupling in the sense of a Holstein-like coupling of the electron charge density with a 
internal degree of freedom of vibration inside the central region, 
and an off-diagonal coupling in the sense of a Su-Schrieffer-Heeger-like coupling \cite{Heeger:1988,Ness:2001} 
to another vibration mode involving the hopping of an electron between 
the central $C$ region and the $L$ electrode. 
A schematic representation of the molecular junction is given in Figure \ref{fig:schema}.

The Hamiltonian for the region $C$ is
\begin{equation}
\label{eq:H_central}
\begin{split}
  H_C 
  = \varepsilon_0 d^\dagger d + \hbar \omega_0 a^\dagger a +
  \gamma_0 (a^\dagger + a) d^\dagger d,
\end{split}
\end{equation}
where $d^\dagger$ ($d$) creates (annihilates) an
electron in the molecular level $\varepsilon_0$. The electron charge
density in the molecular level is coupled to the vibration mode 
of energy $\omega_0$ via the coupling constant $\gamma_0$, and
$a^\dagger$ ($a$) creates (annihilates) a vibration quantum 
in the vibron mode $\omega_0$.
The central region $C$ is nominally connected to two (left and right) one-dimensional
tight-binding chains via the hopping integral $t_{0L}$ and
$t_{0R}$. The corresponding electrode $\alpha=L,R$ self-energy is
$\Sigma^r_\alpha(\omega)=t_{0\alpha}^2 e^{{ i} k_\alpha(\omega)} / \beta_\alpha$
with the dispersion relation 
$\omega=\varepsilon_\alpha+2\beta_\alpha \cos(k_\alpha(\omega))$ where
$\varepsilon_\alpha$ and $\beta_\alpha$ are the tight-binding on-site and off-diagonal
elements of the electrode chains.

To describe the electron-vibron interaction existing across the left contact, we consider
that the hopping integral $t_{0L}$ is actually dependent on some generalised coordinate $X$.
The latter represents either the displacement of the centre-of-mass of the molecule or of
some chemical group at the end of the molecule link to the $L$ electrode. 
At the lowest order, the matrix element can be linearised as
$t_{0L}(X)= t_{0L}+t'_{0L} X$.
Hence the hopping of an electron from the $C$ region to the $L$ region (and {\em vice versa})
is coupled to a vibration mode (of energy $\omega_A$) via the coupling constant $\gamma_A$
(itself related to $t'_{0L}$).
The corresponding Hamiltonian is given by 
\begin{equation}
\label{eq:Vcrossing}
H_{LC} = \gamma_A (b^\dagger + b)(c^\dagger_L d + d^\dagger c_L) + \omega_A b^\dagger b , 
\end{equation}
where $b^\dagger$ ($b$) creates (annihilates) a vibration quantum in the vibron mode $\omega_A$,
the generalised coordinate is $X=\sqrt{{\hbar}/{(2m_A\omega_A)}}(b^\dagger + b)$,
and $c^\dagger_L$ ($c_L$) creates (annihilates) an
electron in the level $\varepsilon_L$ of the $L$ electrode.

The Hamiltonians Eq.~(\ref{eq:H_central}) and Eq.~(\ref{eq:Vcrossing}) are used to calculate 
the corresponding electron self-energies at different orders of the interaction $\gamma_0$ and 
$\gamma_A$ using conventional \NE diagrammatic techniques \cite{Dash:2010,Dash:2011}.

Furthermore, at equilibrium, the whole system has a single and well-defined Fermi 
level $\mu^{\rm eq}$. 
A finite bias $V$, applied across the junction, lifts the Fermi levels as
$\mu_{L,R}=\mu^{\rm eq}+\eta_{L,R} eV$.  The fraction of potential drop \cite{Datta:1997} 
at the left contact is $\eta_L$ and $\eta_R=\eta_L-1$ at the right contact, 
with $\mu_L-\mu_R=eV$ and $\eta_L \in [0,1]$.

\subsection{Self-energies for the interactions}
\label{sec:SE}

In this paper, we consider different approximations to the treat the interaction
inside the central region and the interaction crossing at the left contact.
First of all, calculating exactly the corresponding interaction self-energies is 
a tremendous task since, in principle, they depend on both the phonon progator 
and the electron Green's functions in all the different parts of the system. 
Hence, for the first application of our formalism, we proceed step by step in 
terms of the increasing complexity of the interaction; and we limit ourselves 
to approximations for the self-energies that are well known and well controlled.

The electron-vibron \SEs in the central region $C$ are calculated within the 
self-consistent Born approximation (i.e. diagrams with one phonon line).
The details of these calculations have been reported elsewhere \cite{Dash:2010,Dash:2011}.
For the crossing interaction, we consider a mean-field-like approach for the electron-vibron 
coupling at the $LC$ interface. 
Within such an approximation, we can understand the results of the calculations
in terms of renormalisation of the $LC$ contact.

Furthermore considering the crossing interaction occuring only at one interface 
permit us to check and test the consistency of our formalism. Indeed, with no 
interaction crossing at the right $CR$ interface, the current $I_R$ is given by the 
conventional Meir and Wingreen formula, i.e. Eq.~(\ref{eq:IL_MeirWingreen_bis}) for 
the $CR$ interface. In order to have a consistent formalism, we need to have current 
conservation, and such a constraint is best tested with crossing interaction at only 
one interface. The corresponding results are shown in detail in \ref{sec:Iconserv}.

Within mean-field-like approximations, the effects of the crossing interaction 
correspond to a renormalisation of the coupling in a static or a dynamical manner.
The corresponding self-energies have only retarded and advanced components 
$\Sigma^{r/a}_{LC}$ and $\Sigma^{r/a}_{CL}$. The extra inelastic effects included
in the components $\Sigma^{\gtrless}_{CL}$ are neglected altogether.

Hence the $\Upsilon^X_{C}$ quantities defined in Eq.~(\ref{eq:Upsilons}) become:
\begin{equation}
\label{eq:Upsilons_2}
\begin{split}
\tilde\Upsilon^L_C           & = \Sigma^a_{CL}\ {g}^a_L\ \Sigma^r_{LC} , \\ 
(\tilde\Upsilon^L_C)^\dag      & = \Sigma^a_{CL}\ {g}^r_L\ \Sigma^r_{LC}, \\
\tilde\Upsilon^{L,l}_{C}         & = \Sigma^r_{CL}\ {g}^<_L\ \Sigma^r_{LC}, \\
(\tilde\Upsilon^{L,l}_{C})^\dag  & = - \Sigma^a_{CL}\ {g}^<_L\ \Sigma^a_{LC} .
\end{split}
\end{equation}
with $\Sigma^{r/a}_{LC}=V_{LC}+\Sigma^{\MB,r/a}_{LC}$. 

Correspondingly, the generalised embedding potentials for the left contact 
defined as $Y_C^{L,x} = [\Sigma^r_{CL} g_L \Sigma^r_{LC}]^x$ \cite{Ness:2011,Ness:2012b}
with $x=r,a,\gtrless$,
are now given by
\begin{equation}
\label{eq:YCLx_2}
\begin{split}
Y_C^{L,r}(\omega) = \Sigma^r_{CL} g^r_L \Sigma^r_{LC} , \\
Y_C^{L,a}(\omega) = \Sigma^a_{CL} g^a_L \Sigma^a_{LC} , \\
Y_C^{L,<}(\omega) = \Sigma^r_{CL} g^<_L \Sigma^a_{LC} .
\end{split}
\end{equation}

In the static limit, the mean-field approach leads to the Hartree expression for the 
electron-vibron self-energies at the $LC$ interface:  
\begin{equation}
\label{eq:Vcrossing_Hartree}
\Sigma^{\MB,r/a}_{LC}=- 2\frac{\gamma_A^2}{\omega_A} \langle n_{LC}\rangle ,
\end{equation}
where
\begin{equation}
\label{eq:nLC}
\langle n_{LC}\rangle = - i \int\frac{{\rm d}\omega}{2\pi}\ G^<_{LC}(\omega) .
\end{equation}

The self-energy $\Sigma^{\MB,r/a}_{LC}$ is independent of the energy $\omega$
and leads to a static 
(nonetheless bias-dependent) renormalisation of the nominal coupling $V_{CL}=V_{LC}=t_{0L}$ 
between the $L$ and $C$ regions. 
This NE renormalisation will induce, among other effects, a bias-dependent modification 
of the broadening of the spectral features of the $C$ region.
We have analysed these effects in details in Ref.\cite{Ness:2012}. The
renormalisation is such that the amplitude of the current 
is reduced in comparison with the current values obtained 
when the interaction is present only in the central region. 
The NE static renormalisation of the contact is highly
non-linear and non-monotonic in function of the applied
bias, and the larger effects occur at applied biases corresponding 
to resonance peaks in the dynamical conductance. 
The conductance is also affected by the NE renormalisation of the 
contact, showing asymmetric broadening around the resonance peaks and 
some slight displacement of the peaks at large bias in function of the 
coupling strengh $\gamma_A$.

Beyond the static limit, we can develop a dynamical mean-field-like approach for the
electron-vibron coupling following Ref.~\cite{Ciuchi_S:1997}.
The retarded self-energy containing all orders
of the electron-vibron coupling is expressed as a continued fraction as 
shown analytically in Refs.~\cite{Ness:2006,Cini:1988}. We have already used
such an approach to study the transport properties of organic molecular wires
which are dominated by the propagation of polarons \cite{Ness:1999,Ness:2001}
or solitons \cite{Ness:2002a}.
The expression for the corresponding self-energy $\Sigma^r_{LC}(\omega)$ is
given by:
\begin{equation}
  \label{eq:SEr_contfrac}
  \begin{split}
   \Sigma^r_{LC}(\omega)  =
    \cfrac{\gamma_A^2}{
      G^r_{LC}(\omega-\omega_A)^{-1}-
      \cfrac{2\gamma_A^2}{
        G^r_{LC}(\omega-2\omega_A)^{-1}-
        \cfrac{3\gamma_A^2}{
          G^r_{LC}(\omega-3\omega_A)^{-1}- ...}}} 
  \end{split}
\end{equation}
where $G^r_{LC}$ is the retarded component of the off-diagonal GF between the central
region and the left lead $L$. Its closed expression is obtained from the corresponding
Dyson equation $G^r_{LC}=[g\Sigma G]^r_{LC}$:
\begin{equation}
\label{eq:GrLC_bis}
G^r_{LC}(\omega) =
\left[
[g^r_L]^{-1} - 
\Sigma^r_{LC}
\left[ [\tilde g^r_C]^{-1} - Y_C^{R,r} \right]^{-1}
\Sigma^r_{CL}
\right]^{-1} 
\Sigma^r_{LC}
\left[
[\tilde g^r_C]^{-1} - Y_C^{R,r}
\right]^{-1} ,
\end{equation}
with
$\tilde g^r_C(\omega) = [\omega - \varepsilon_0 - \Sigma_C^{\rm MB,r}(\omega) ]^{-1}$ 
and 
$Y_C^{R,r}(\omega) = V_{CR} g^r_R(\omega) V_{RC}$.

Such a dynamical mean-field-like approach, which corresponds to a quasi-particle approach
since the self-energy $\Sigma^{r/a}_{LC/CL}$ can be incorporated into a Schr\"odinger-like
equation, is expected to affect the transport properties via the non-equilibrium dynamical
renormalisation of the contacts, i.e. through the generalised embedding potentials
$Y_C^{L,x}(\omega)$ as well as through the corresponding $\Upsilon^x_{C}(\omega)$ quantities.

Finally, note that for the lowest-order expansion, the self-energy $\Sigma^r_{LC}$ takes
a simple form:
\begin{equation}
  \label{eq:SEr_lowest}
  \Sigma^r_{LC}(\omega)  = {\gamma_A^2} G^r_{LC}(\omega-\omega_A) \ .
\end{equation}

\section{Results}
\label{sec:res}

We have perfomed calculations, in a self-consistent manner, for many different values 
of the Hamiltonian parameters. 
We present below the most characteristic results for
different transport regimes and for different coupling strengths $\gamma_A$,
while the interaction in the region $C$ is taken to be in the intermediate 
coupling regime $\omega_0 = 0.2$, $\gamma_0 = 0.15$, i.e. $\gamma_0/\omega_0 = 0.75$.
 
The nominal couplings between the central region and the electrodes $t_{0L,R}$, 
before NE renormalisation, are not too large, so that we can discriminate clearly 
between the different vibron side-band peaks in the spectral functions.
The values chosen for the parameters are typical values for realistic molecular 
junctions \cite{Dash:2012,Dash:2011}.
In the following the current is given in units of charge per time, the conductance
in units of the quantum of conductance $G_0=2e^2/\hbar$ and the bias $V$ and the 
embedding potential $Y_C$ in natural units of energy where $e=1$ and $\hbar=1$.

\subsection{Conserving approximation}
\label{sec:Iconserv}

One of the most important physical conditions that our formalism needs to 
fulfil is the constraint of current conservation. We use conserving approximations
to calculate the interaction self-energies in the central region $C$ and an quasi-particle
approximation for crossing interaction at the left interface.

\begin{figure}
  \includegraphics[clip,width=\textwidth]{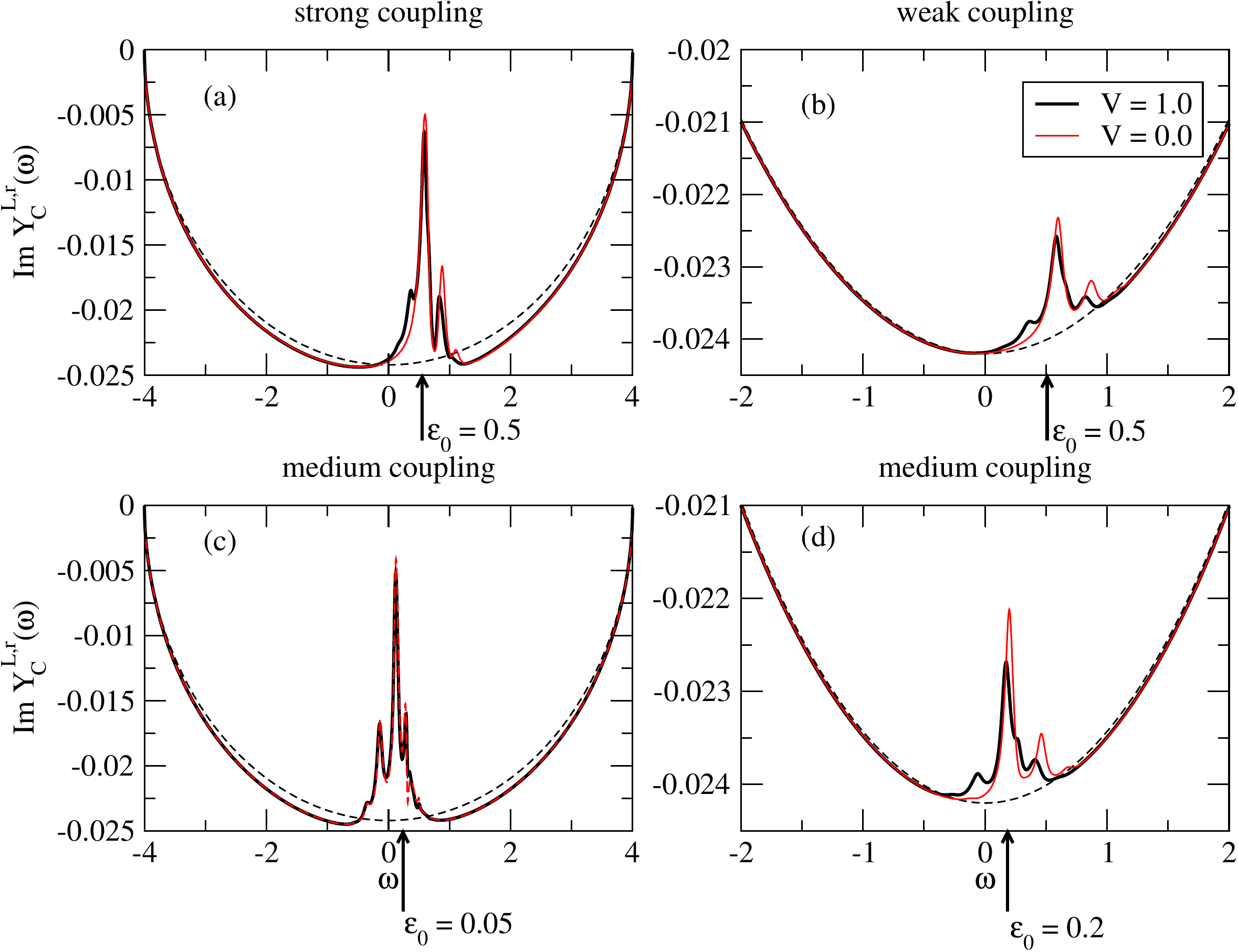}
  \caption{(Color online) Imaginary part of the left-lead generalised embedding potential
  $Y_C^{L,r}(\omega)$. 
  The dashed lines represent the $L$ lead spectral function in the absence of renormalisation
  at the $LC$ interface (i.e. semi-elliptic density of states). The thin (red) lines are the
  spectral functions $\Im m Y_C^{L,r}(\omega)$ at equilibrium ($V=0$). The thick (black) 
  lines are the non-equilibrium spectral functions $\Im m Y_C^{L,r}(\omega)$ at finite bias
  ($V=1.0$). These show a strong reduction of the spectral functions on energy range of
  $\Delta\omega \approx 1$ around  $\varepsilon_0$.
  Panel (a) Off-resonant regime $\varepsilon_0=0.5$ and strong coupling at the interface
  $\omega_A=0.20$, $\gamma_0=0.28$.
  (b) Off-resonant regime $\varepsilon_0=0.5$ and weak coupling at the interface
  $\omega_A=0.20$, $\gamma_0=0.07$.
  (c) Resonant regime $\varepsilon_0=0.05$ and medium coupling at the interface
  $\omega_A=0.20$, $\gamma_0=0.14$.
  (d) Intermediate regime $\varepsilon_0=0.2$ and medium coupling at the interface
  $\omega_A=0.10$, $\gamma_0=0.07$.
  The other parameters are $\omega_0=0.20$, $\gamma_0=0.15$, $t_{0R}=t_{0L}=0.22$, 
  $\beta_\alpha=2.0, \epsilon_\alpha=0.0$, $\eta_V = 1$.}
  \label{fig:ImYleft}
\end{figure}

With our choice for $\Sigma^r_{LC}(\omega)$ given by Eq.~(\ref{eq:SEr_contfrac}), 
we find that the left lead generalised embedding potential $Y_C^{L,r}(\omega)$ is 
renormalised by the crossing interaction. This is clearly seen in Figure \ref{fig:ImYleft},
which shows the imaginary part of  $Y_C^{L,r}$ for different transport regimes,
different coupling at the $LC$ interface and for both equilibrium and non-equilibrium
conditions.

At equilibrium, and in the absence of contact renormalisation, the imaginary part
of $Y_C^{L,r}$ is simply the imaginary part of the conventional $L$ lead self-energy
$\Sigma^r_L(\omega)=t_{0L}^2 e^{{ i} k_L(\omega)} / \beta_L$ and corresponds to
a semi-elliptic functions with non-zero values within the energy range
$-2\beta_L \le \omega \le +2\beta_L$
(see dashed lines in figure \ref{fig:ImYleft}).
The spectral weight of $\Im m\ \Sigma^r_L$ is given by
$\int {\rm d}\omega\ \Im m\ \Sigma^r_L(\omega) 
= t_{0L}^2 \int {\rm d}\omega\ \Im m\ g^r_L(\omega) 
= - \pi t_{0L}^2$. 

In the presence of the renormalised $L$ contact, we obtained a strong deviation from the
semi-elliptic function as shown by the (red) thin lines and (black) thick lines
in Figure \ref{fig:ImYleft}, which correspond respectively to the equilibrium (bias
$V=0$) and the non-equilibrium (bias $V=1.0$) conditions. This result indicates
a strong reduction of the available transport channels in the renormalised $L$ lead 
embedding potential for regions of energy where $\Sigma^{r/a}_{LC/LC}(\omega)$ has 
non-zero value.

However, a very important property is that the total spectral weight of 
$\Im m\ \Sigma^r_L$ is conserved for all the calculations we have performed, i.e.
for all the different transport regimes, coupling strength $\gamma_A$ and $\omega_A$
at the $LC$ interface and all applied bias.
We find that in all the cases $\int {\rm d}\omega\ \Im m\ Y_C^{L,r}(\omega) = - \pi t_{0L}^2$
(with maximum deviation of $0.05 \%$ arising from numerical errors in the results
shown in Figure \ref{fig:ImYleft}). So in this sense, we can say that the approximation
for $\Sigma^{r,a}_{LC/LC}$ is conserving.

Correspondingly, we have also carefully checked that the current is conserved for all 
the calculations presented in the present paper, i.e. that 
$\vert I_L+I_R \vert / \vert I_L \vert \sim \vert I_L+I_R \vert / \vert I_R \vert < 10^{-5}$,
with the current $I_L$ at the $LC$ interface is given by expression Eq.~(\ref{eq:ILfinal})
and the current $I_R$ at the $CR$ interface does not contain any contact renormalisation,
and hence is given by a Meir and Wingreen like expression Eq.~(\ref{eq:IL_MeirWingreen_bis}).

\subsection{Dynamical non-equilibrium renormalisation}
\label{sec:dynNErenorm}

The dynamical renormalisation of the $L$ lead embedding potential $Y_C^{L,r}$
presents features (dips) at some energies. Qualitatively speaking, there is a
form of correlation between them and the features that exist in the spectral 
function of the central region $A_C(\omega)=(G^a_C(\omega)-G^r_C(\omega)) / 2\pi i$.

\begin{figure}
  \includegraphics[clip,width=\textwidth]{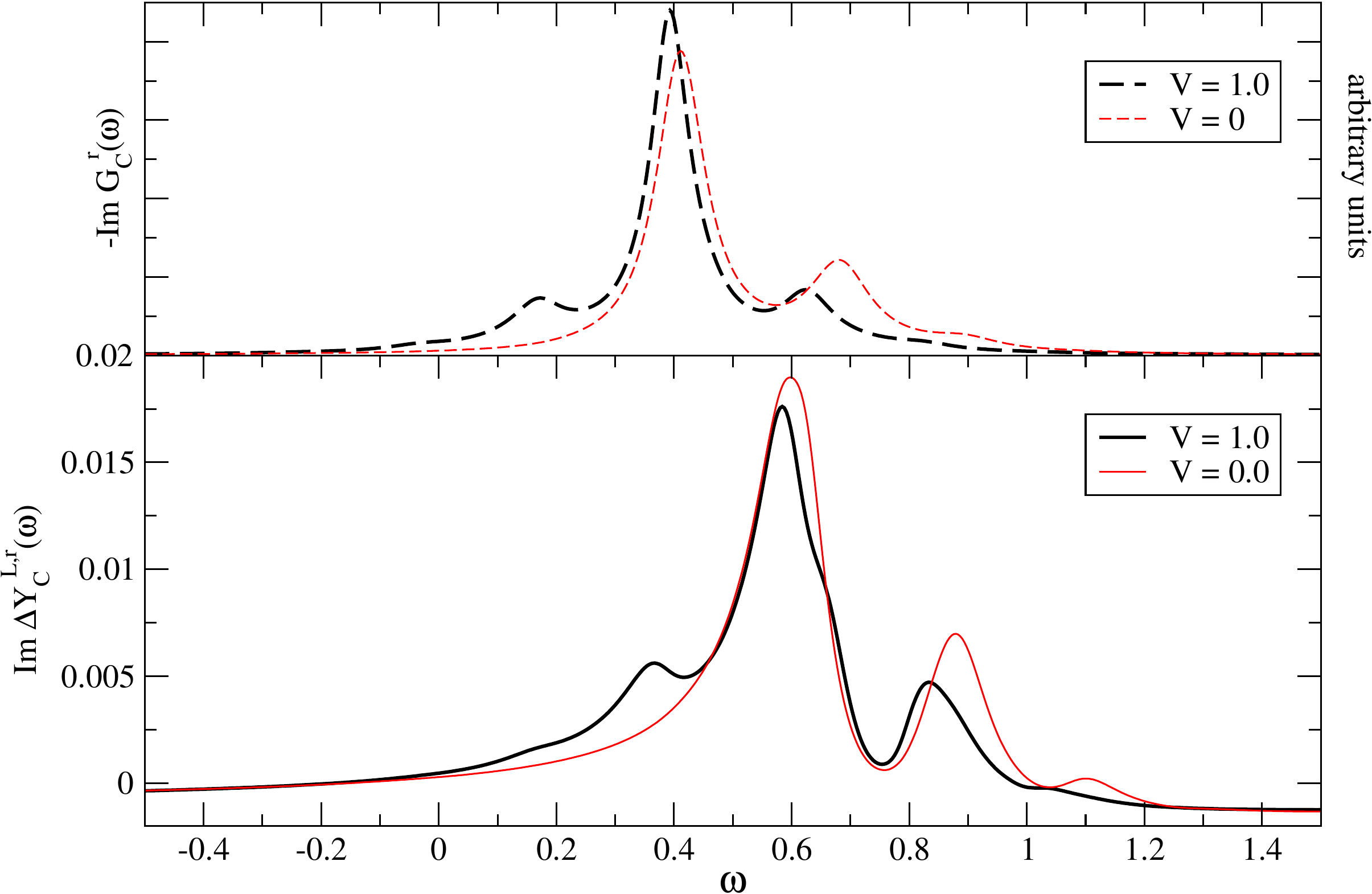}
  \caption{(Color online) Rescaled spectral function proportional to $-\Im m G^r_C$
  for the central region (top panel) and imaginary part of the $L$ lead generalised embedding
  potential relative to the non-renormalised $L$ lead self-energy:
  $\Delta Y_C^{L,r}=Y_C^{L,r}-\Sigma_L^r$ (bottom panel).
  The calculations are performed for the same parameters as in panel (a) of
  Figure \ref{fig:ImYleft} and for both zero bias and finite bias $V=1$. Here $\mu_R=0$
  and $\mu_L=V$.
  The qualitative correlations between the features in $\Delta Y_C^{L,r}$ and in the spectral 
  function of the $C$ 
  region $-\Im m G^r_C$ are clearly observed.}
\label{fig:AwC_renormSigmallead}
\end{figure}

Figure \ref{fig:AwC_renormSigmallead} shows the spectral function $A_C(\omega)$,
rescaled by some factor, for a set of parameters corresponding to panel (a)
in figure \ref{fig:ImYleft}. The bottom panel of figure \ref{fig:AwC_renormSigmallead}
shows the imaginary part of $Y_C^{L,r}(\omega)$ from which the semi-elliptic background
has been subtracted, i.e. $\Delta Y_C^{L,r}=Y_C^{L,r}-\Sigma_L^r$. Qualitatively 
speaking, both quantities present similar features, shifted in energy by an amount
corresponding to $\omega_A$ ($\omega_A=0.2$ in the calculations) as expected from
the definition of $\Sigma^{r/a}_{LC/LC}(\omega)$.

One can then expect a maximum effect of the dynamical renormalisation in an integrated
quantity such as the current when the features in $Y_C^{L,r}(\omega)$ coincide with
those in the spectral function $A_C(\omega)$ within a given bias energy-window.

\begin{figure}
  \includegraphics[clip,width=0.55\textwidth]{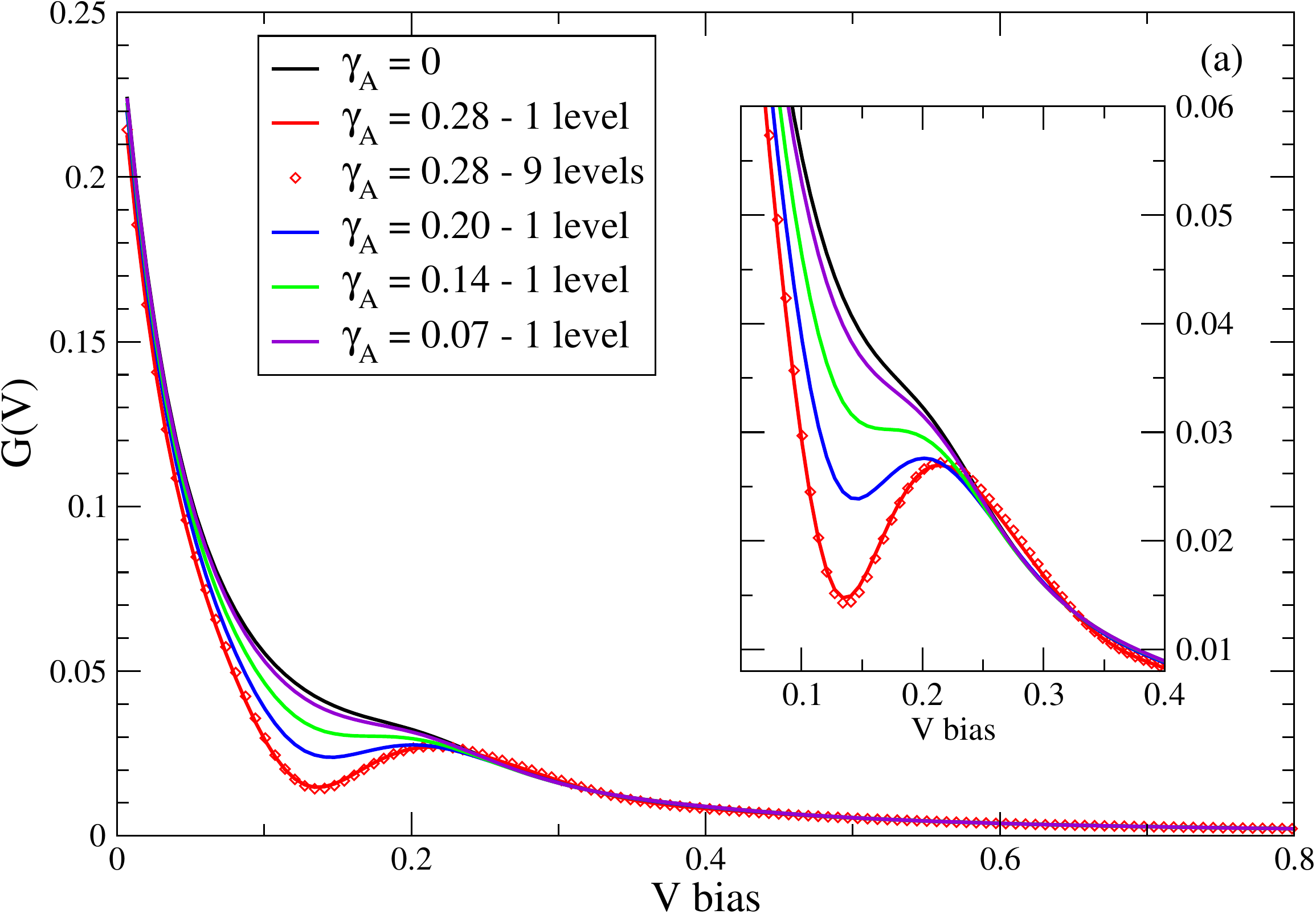}
  \includegraphics[clip,width=0.55\textwidth]{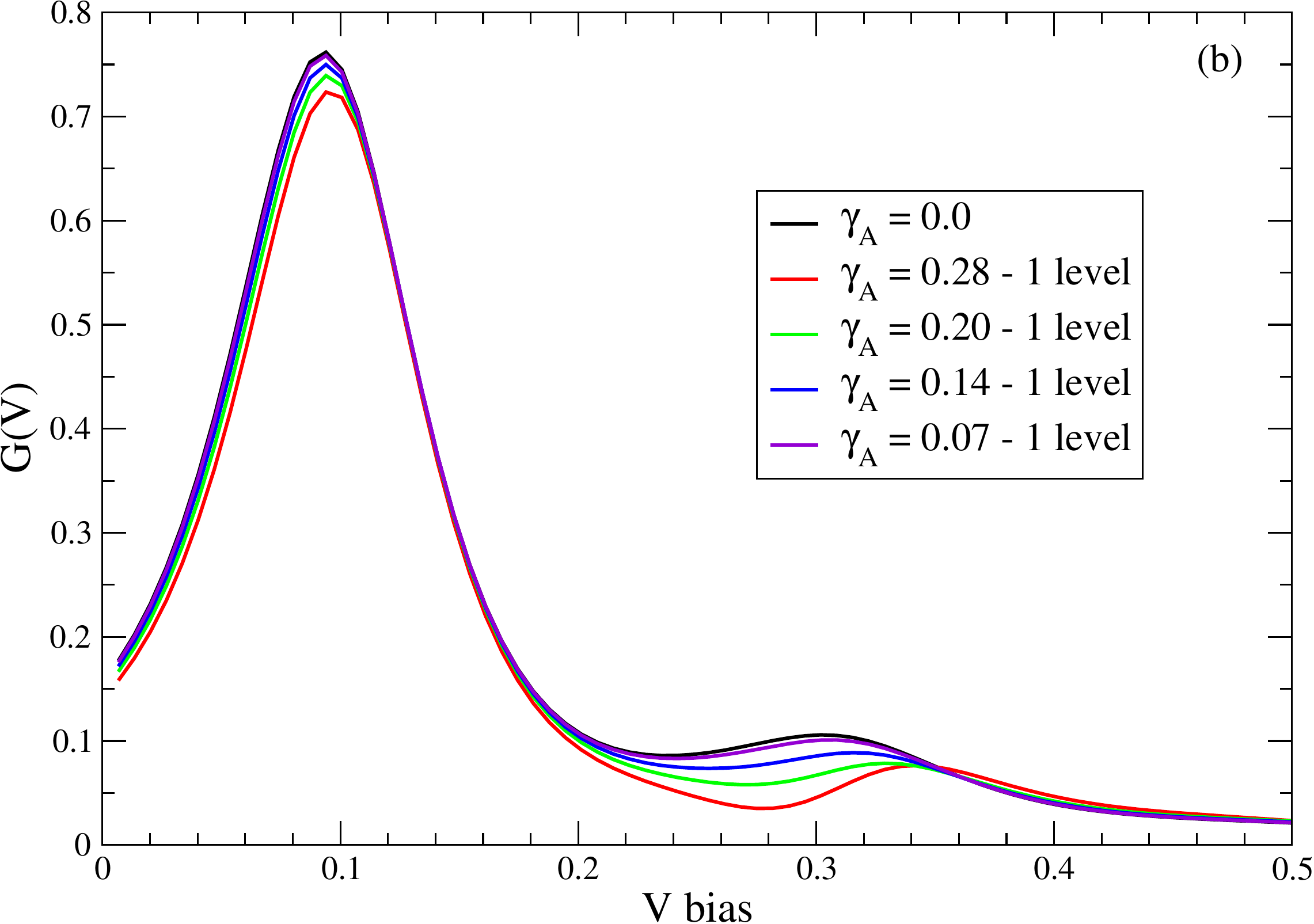}

  \includegraphics[clip,width=0.55\textwidth]{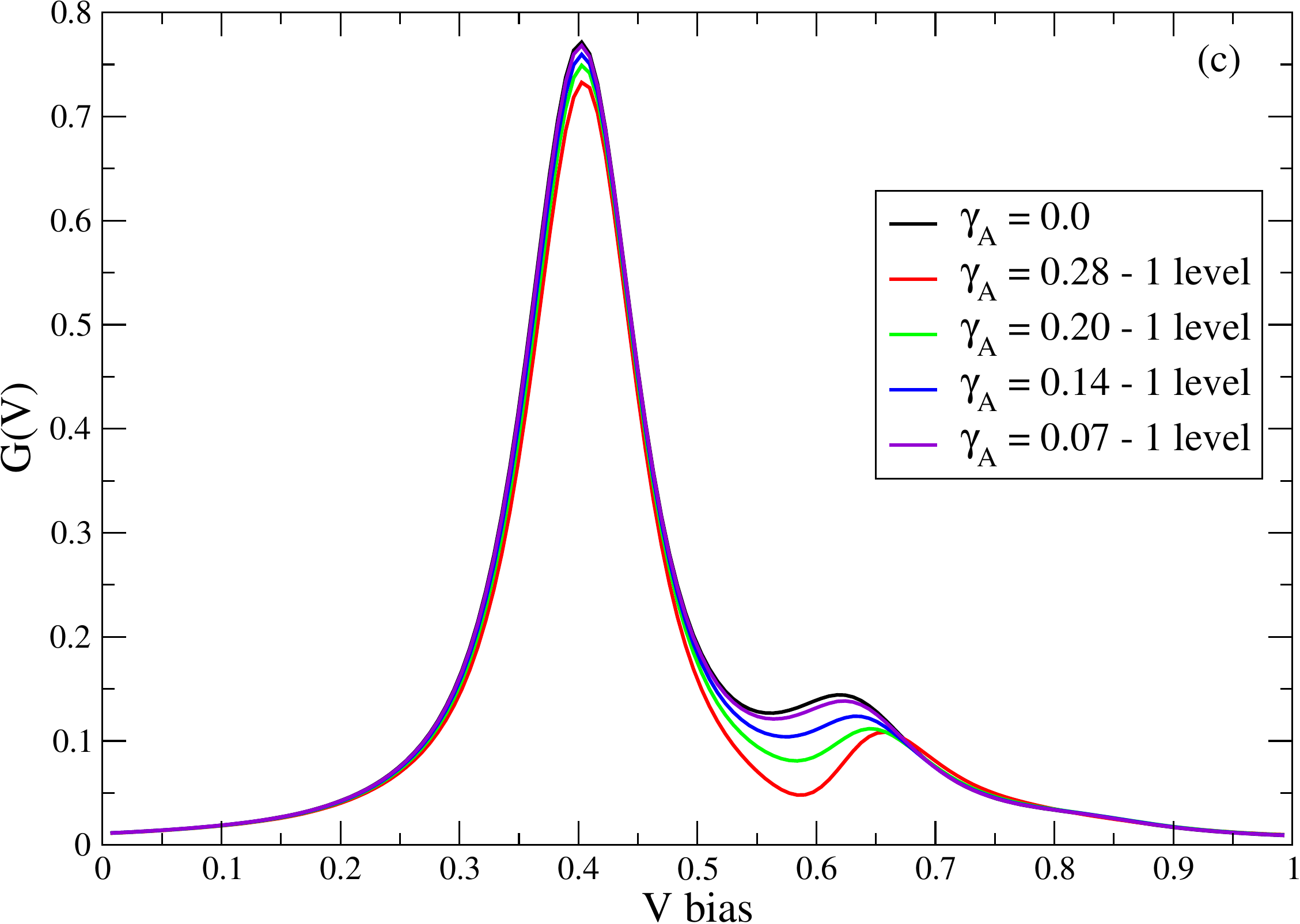}

  \caption{(Color online) Dynamical conductance $G(V)$ for different transport regimes
  and different coupling strength $\gamma_A$, and $\omega_A=0.20$.
  Panel (a) Resonant regime $\varepsilon_0=0.05$.
  Panel (b) Off-resonant regime $\varepsilon_0=0.2$.
  Panel (c) Off-resonant regime $\varepsilon_0=0.5$.
  The non-equilibrium dynamical renormalisation of the $LC$ contact affects both the
  main conductance peak and the first vibron side-band peak.
  The other parameters are $\omega_0=0.20$, $\gamma_0=0.15$, 
  $t_{0R}=t_{0L}=0.22$, 
  $\beta_\alpha=2.0, \epsilon_\alpha=0.0$, $\eta_V = 1$.}
\label{fig:GV_omega020_vsGammaA}
\end{figure}

\begin{figure}
  \includegraphics[clip,width=0.55\textwidth]{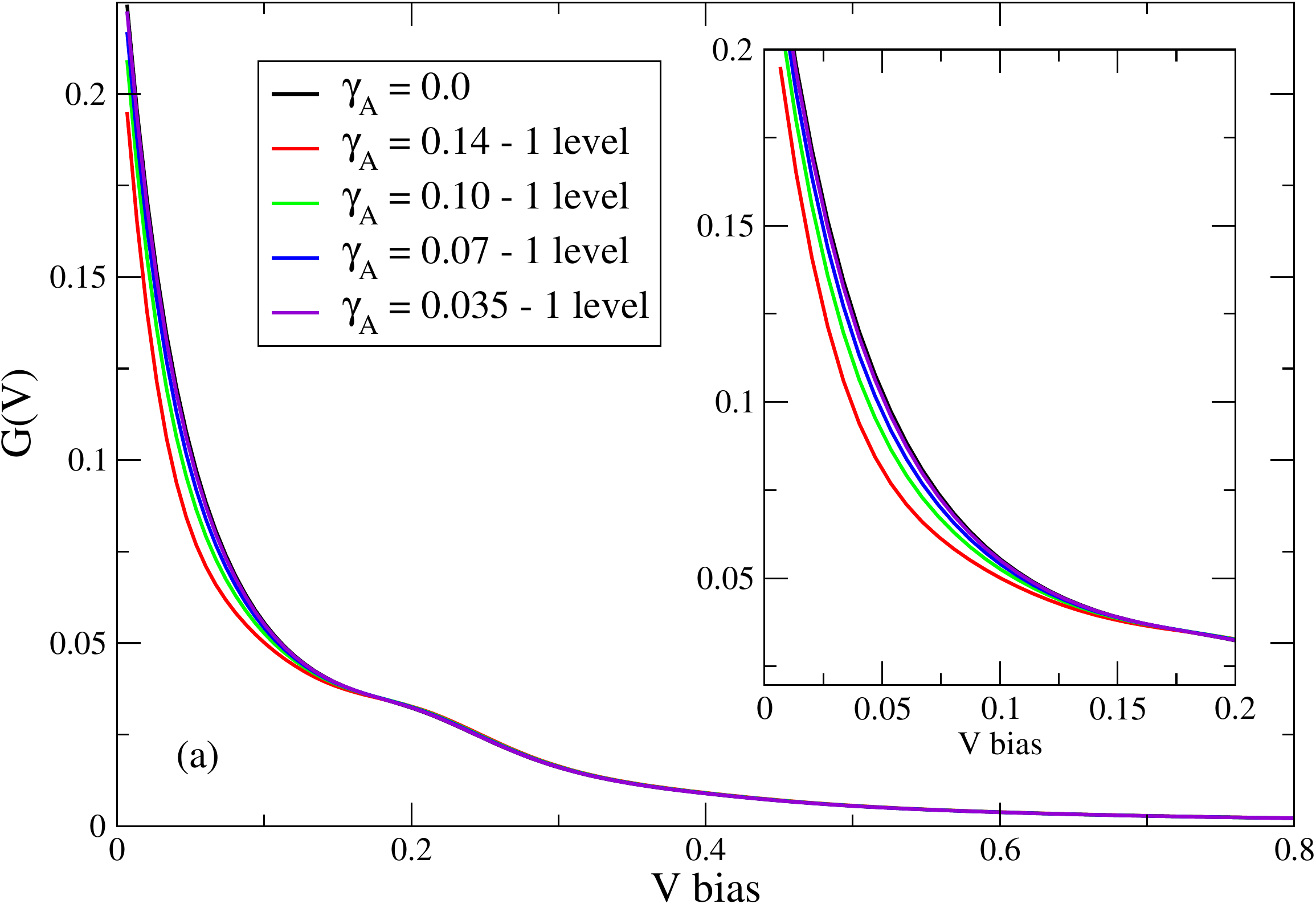}
  \includegraphics[clip,width=0.55\textwidth]{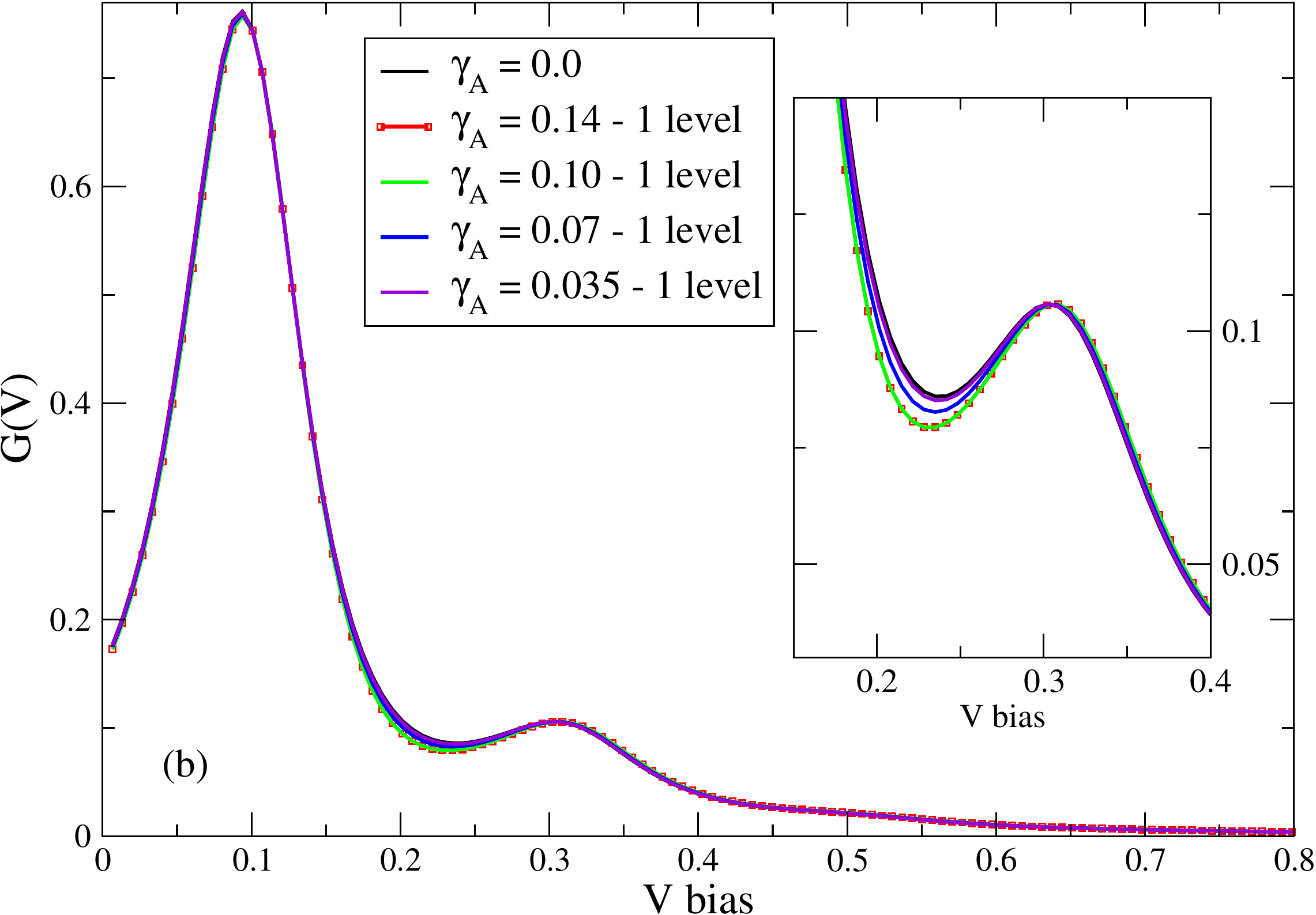}

  \includegraphics[clip,width=0.55\textwidth]{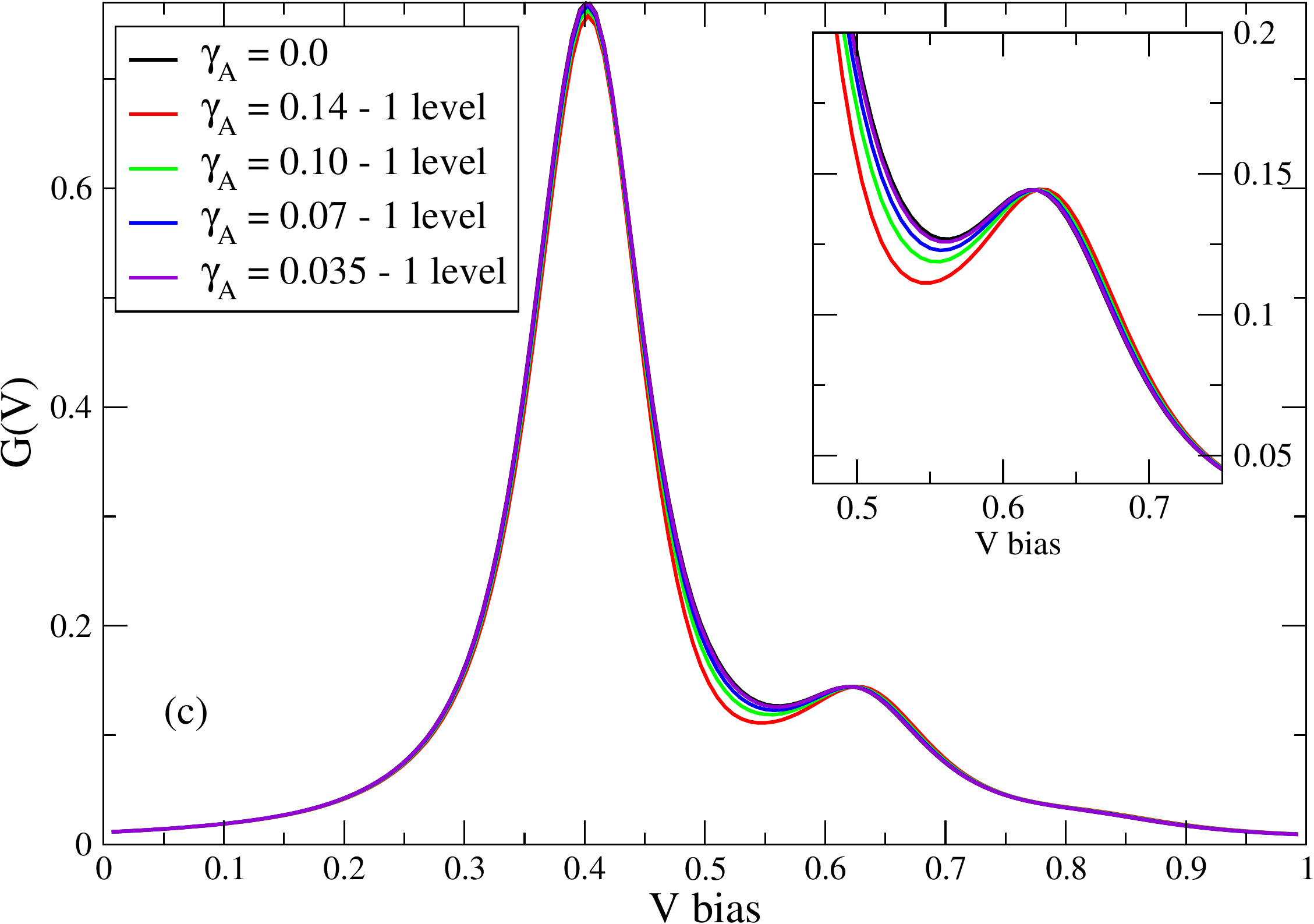}

  \caption{(Color online) Dynamical conductance $G(V)$ for different transport regimes
  and different coupling strength $\gamma_A$, and $\omega_A=0.10$.
  Panel (a) Resonant regime $\varepsilon_0=0.05$.
  Panel (b) Off-resonant regime $\varepsilon_0=0.2$.
  Panel (c) Off-resonant regime $\varepsilon_0=0.5$.
  The non-equilibrium dynamical renormalisation of the $LC$ contact affects both the
  main conductance peak and the first vibron side-band peak.
  The other parameters are $\omega_0=0.20$, $\gamma_0=0.15$, 
  $t_{0R}=t_{0L}=0.22$, 
  $\beta_\alpha=2.0, \epsilon_\alpha=0.0$, $\eta_V = 1$.}
\label{fig:GV_omega010_vsGammaA}
\end{figure}

We now consider the modification of the dynamical conductance $G(V)=dI/dV$ induced by
the crossing interaction $\Sigma^{r/a}_{LC/LC}(\omega)$ for different transport 
regimes and for different crossing interaction strength.
Figures~\ref{fig:GV_omega020_vsGammaA} and \ref{fig:GV_omega010_vsGammaA} show $G(V)$
for three different transport regimes for weak to strong crossing interaction strength
$\gamma_A$, and for two values of $\omega_A$ ($\omega_A=0.2$ for Fig.~\ref{fig:GV_omega020_vsGammaA}
and $\omega_A=0.1$ Fig.~\ref{fig:GV_omega010_vsGammaA}).

The dynamical renormalisation of the left contact slightly affects the main resonance
peak in the conductance, and more importantly the lineshape of the first vibron side-band
peak above the main conductance peak.
The most important effects are obtained for the strong crossing interaction strength 
$\gamma_A=0.28$. Even if the ratio $\gamma_A/\omega_A$ is conserved between the calculations
shown in Figure~\ref{fig:GV_omega020_vsGammaA} and Figure~\ref{fig:GV_omega010_vsGammaA},
the absolute value of $\gamma_A$ is the crucial quantity that governs the effects of the
dynamical renormalisation.
Furthermore, for the different sets of parameters we have considered, our calculations
show that the lower order expansion for $\Sigma^{r}_{LC}$ (see Eq.~(\label{eq:SEr_lowest})) 
provides a good
approximation to the results obtained with a larger number of levels in the continued
fraction expansion of  $\Sigma^{r}_{LC}$ (see panel (a) in fig.~\label{fig:GV_omega020_vsGammaA}).

\begin{figure}
  \includegraphics[clip,width=0.55\textwidth]{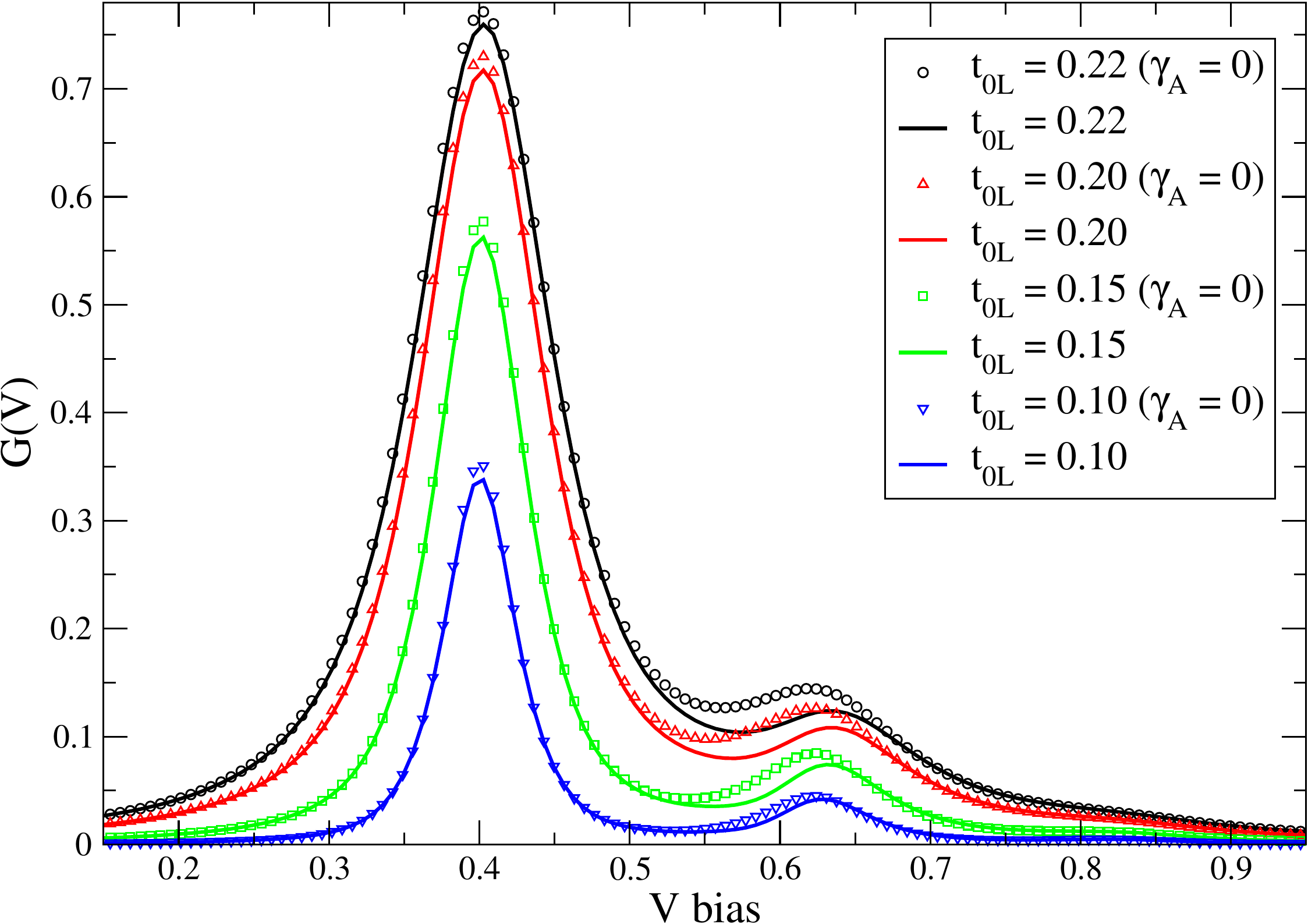}
  \includegraphics[clip,width=0.55\textwidth]{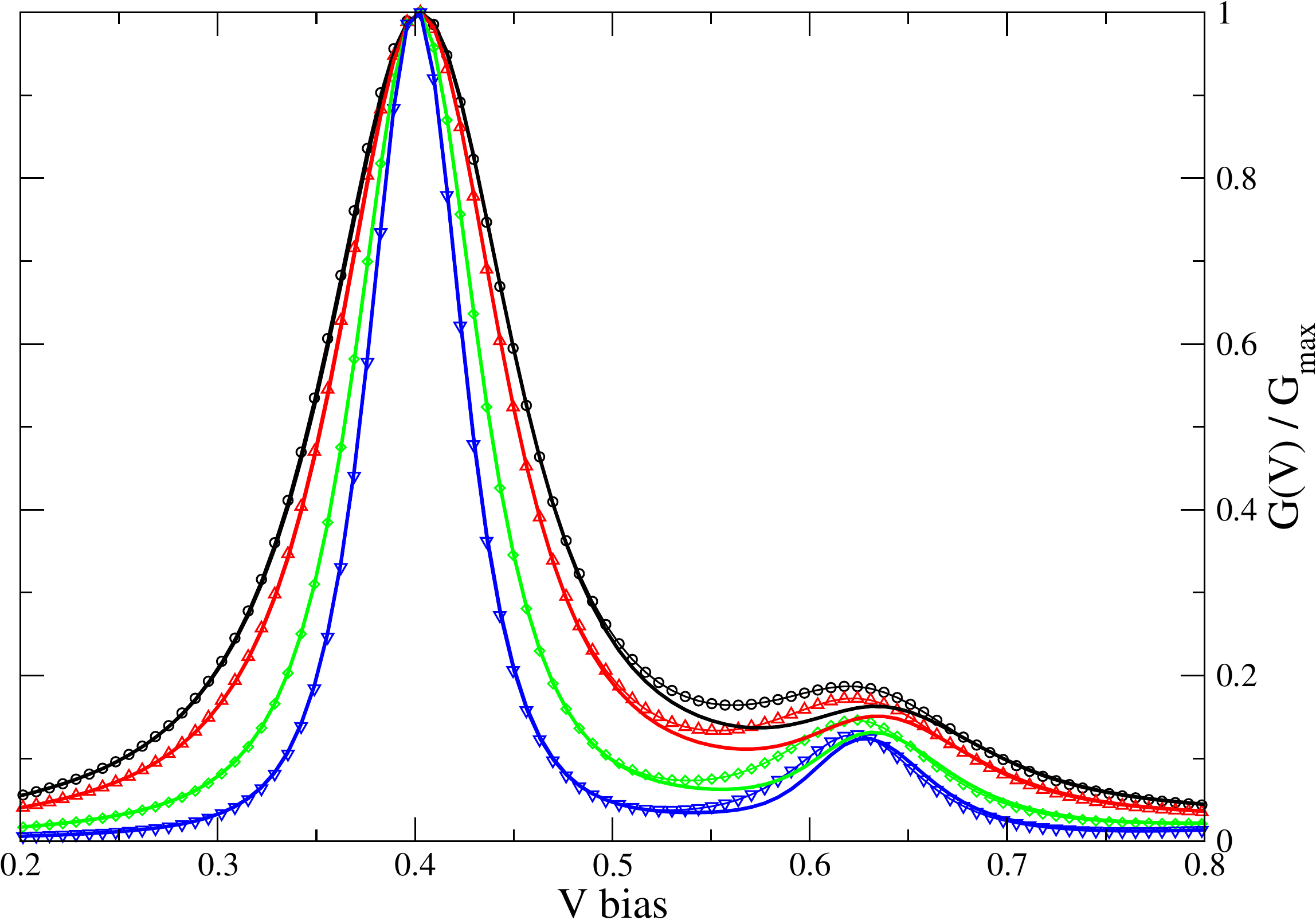}
  \caption{(Color online) (Left Panel) Dynamical conductance $G(V)$ for the off-resonant transport regime
  ($\varepsilon_0=0.50$) and different norminal coupling strength $t_{0R}=t_{0L}$ to the leads.
  (Right Panel) Dynamical conductance $G(V)$ rescaled such that the amplitude of the main resonance
  peak is 1 quantum of conductance.
  The other parameters are $\gamma_A=0.2$, $\gamma_A=0.14$ (unless otherwise indicated), $\omega_0=0.20$, 
  $\gamma_0=0.15$, 
  $\beta_\alpha=2.0, \epsilon_\alpha=0.0$, $\eta_V = 1$.}
\label{fig:GV_omega020_vst0LR}
\end{figure}
 
We now check the effects of varying the nominal coupling $t_{0\alpha}$ on the conductance. 
The results are shown in Figure~\ref{fig:GV_omega020_vst0LR} for the off-resonant transport
regime with and without dynamical renormalisation. 
With renormalisation of the left contact, the conductance values decrease with increasing coupling
$t_{0\alpha}$ to the leads while the conductance peaks broaden. This is seen more clearly 
in the right panel of Figure~\ref{fig:GV_omega020_vst0LR} where the rescale conductance $G(V)/G_{\max}$
is plotted such that the main resonance peak has an amplitude of 1. 

In the presence of dynamical renormalisation of the $L$ contact, the main conductance peak follows the
same behaviour as in the absence of renormalisation. However, one observes a highly non-linear behaviour
of the modification of the first vibron side-band peak.
The complete and detailed understanding of such modification is rather complex, and beyond the scope
of the present paper. However it is strongly related to the new features in the renormalised $L$ lead
embedding potential which depart from the non-interacting case.

\section{Conclusion}
\label{sec:ccl}

We have studied the transport properties through a two-terminal nanoscale device with interactions
present not only in the central region but also with interaction crossing at the interface between
the left lead and the central region. 
To calculate the current for such a fully-interacting system, we have used our
recently developed quantum transport formula \cite{Ness:2011} based on the NEGF formalism.
As a first practical application, we have considered a prototypical single-molecule nanojunction
with electron-vibron interaction. In terms of the electron density matrix, the interaction is diagonal 
in the central region for the first vibron mode and off-diagonal between the central region
and the left electrode for the second vibron mode.
The interaction self-energies are calculated in a self-consistent manner using the lowest order
Hartree-Fock-like diagram in the central region and a quasi-particle (dynamical mean-field-like)
approach for the crossing interaction.
Our calculations were performed for different transport regimes ranging from the far off-resonance
to the quasi-resonant regime, and for a wide range of parameter values.

They show that, for this model, we obtain a non-equilibrium (i.e. bias dependent)
dynamical (i.e. energy dependent) renormalisation of the generalised embedding potential of
the $L$ left lead. The renormalisation is such that the amplitude of the corresponding
spectral function is reduced around the molecular level energy $\varepsilon_0$ over an
energy range roughly equal to the energy support of the spectral density of the central
region $C$.
This corresponds a reduction of the number of transport channels in the left lead at
given energy, even if the total spectral weight of the $L$ embedding potential is conserved.

The non-equilibrium dynamical renormalisation of the $L$ contact is highly non-linear and 
non-monotonic in function of the applied bias, and the larger effects occur at applied bias 
for which features are present in both the spectral function of the central region $C$ and
the generalised embedding potential of the $L$ lead. 
The conductance is affected by the NE renormalisation of the contact: the amplitude of the
main resonance peak is modified as well as  the lineshape of the first vibron side-band peak.

Finally, extensions of the present study are now considered to go beyond the quasi-particle 
approach for the crossing interaction self-energy by including as well as the lesser and 
greater components of the $LC$ interface self-energy.
This should lead to dynamical NE renormalisation of the contact involving inelastic scattering 
processes.

\section*{References}

\providecommand{\newblock}{}

\end{document}